\documentclass[doublecol,linenumbers]{epl2} 
\title{Strain designed Josephson $\pi$ junction qubits with topological insulators} 
\author{Colin Benjamin} %\email{cbiop@yahoo.com}
\institute{ National institute of Science education \& Research, Bhubaneswar 751005, India }
\abstract { A Josephson qubit is designed via the application of a tensile strain to a topological insulator surface sandwiched between two s-wave superconductors. The strain applied leads to a shift in the Dirac point without changing the pre-existing conducting states, on the surface of a topological insulator. Strain applied can be tuned to form a $\pi$ junction in such a structure. Combining two such junctions in a ring architecture leads to the ground state of the ring being in doubly degenerate state- the "0" and "1" states of a qubit. A qubit designed this way is quite easily controlled via the tunable strain applied. We report on the conditions necessary to design such a qubit. Finally the operating time of a single qubit phase gate is derived.}
\date{\today}
\begin{document}

\maketitle
  \section { Introduction} 
In our continuing quest for smaller and faster computers with large storage capacity we would soon be breaking into the quantum limit. This holds enormous challenges as well as opportunities. In this letter  a novel material called topological insulator(TI) is investigated\cite{zhang}. It is a novel material with fascinating mechanical and electrical properties among which are faster electronic speeds, large tensile strength and most intriguingly the physics governing these materials is the same as that of particles at the Large Hadron Collider in CERN, Geneva namely the Dirac equation in contrast to the Schrodinger equation which forms the basis for our understanding of almost all other materials. This makes it not only of fundamental importance to material scientists but also to theoretical physicists trying to understand the origins of the universe and significantly to computer hardware developers trying to build a computer which works according to quantum principles\cite{moore}. Quantum computers (QC's)  offer the prospect of massive parallel processing since qubits (quantum equivalent of the bit) can be used in more than one calculation at a time\cite{vedral}. The aim of this letter is to theoretically propose ways and means to design the basic components of a QC- qubits  and gates using nothing more than a mechanical strain applied to a TI\cite{zhang, moore} sandwiched between two superconductors.  A strain engineered  layer of TI can act as a template for an all integrated nanocircuit\cite{jinfeng,zhao}. Most uniquely strain has been shown experimentally to be tunable which implies a control exclusive of any magnetic or electric fields, the usual modes of control in almost all qubit proposals. This is more lucrative since at the nanoscale these fields are most unwieldy as they directly couple to qubit states- making them fragile (losing their quantum properties in a very short time to a process called decoherence). Conversely, strain control is indirect since it affects only the material (how atoms attach to each other). {In this letter, strain is the reason for the provenance of the qubit and the qubit is controlled by it too, this is unlike other proposals which aim to use it only as an external control- the qubit originates due to a different process altogether\cite{lovett}}.

\begin{figure}\centerline{\includegraphics[width=9cm,height=10cm]{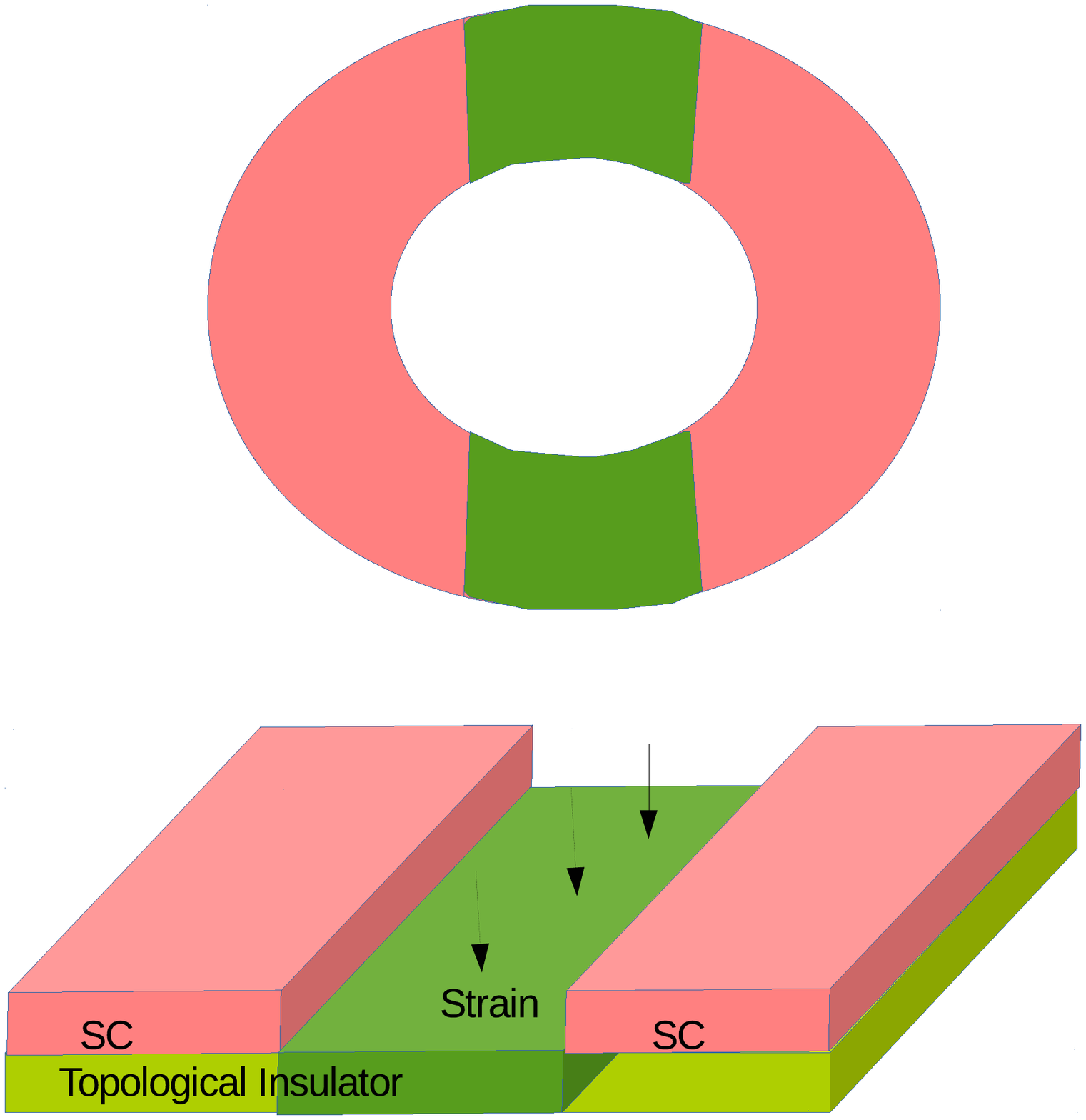}} \vskip -3cm \caption{A. Two semi-circular superconducting strips(colored pink) with TI layers(colored green) on top and bottom. Application of a strain to the TI layers leads to a $0$ or $\pi$ phase shift in the current and ground state. B. Topological insulator Josephson junction(JJ) with strained layer in between.  \label{pi-new}} \end{figure} 
TI's such as $Bi_{2}Se_{3}$ and $HgTe$ are new quantum states of matter with an insulating bulk and topologically protected conducting surface states with a single Dirac cone at the $\Gamma$ point\cite{kane}.  This is not unlike graphene with the exception that graphene has two Dirac cones\cite{katsnelson}.  The conducting states on surface of a TI are extremely robust against any perturbations such as point defects and impurities\cite{zhang}. Recent\cite{jinfeng,yingliu} experiments conducted on TI's have shown that these topological surface states can be manipulated or suppressed by means of strain. However, strain is by no means the only way to manipulate the topological surface states on a TI. Doping the bulk in these materials can also change these states, however this change is permanent, once altered the states don't revert back to their original likeness.  Strain designed manipulations of the surface states on the other hand are reversible and therefore a great way to bring about external control over these states. By external control we can control the amount of current flowing in these states and thereby design quantum switches and gates.  But not any kind of  strain will do, only a tensile strain helps us perform  these manipulations. A compressive strain completely destroys these Dirac states and then they are of no use to us. We in this work predict that a tensile strain applied to a TI sandwiched between two s-wave superconductors can give rise to a $\pi$-junction, which has its ground state at a phase difference $\pi$ unlike the usually observed JJ's with ground state at $0$. The phase difference referred to here is that between the macroscopic phases of the superconductors on either side of the TI. The advantage of a TI  based $\pi$-junction over other such junctions is that in the latter the $\pi$-phase shift is difficult to manipulate, while in Dirac materials like TI it isn't so because of the ease at which even a small gate voltage can tune the Fermi energy. Combining two such junctions we can design a qubit.  Qubits based on Dirac materials, notably graphene, have already been theoretically predicted\cite{trauze,benj-pi}.  In this proposal the aim will be to go beyond the spin\cite{trauze} and d-wave  $\pi$-junction qubits\cite{benj-pi}.{The goal will be to exploit the  $\pi$-shift seen in JJ not by using d-wave superconducting correlations but via a tensile strain with normal s-wave superconductors.}  However instead of graphene we will be using TI's because $\pi$-junction is not seen when a strained graphene layer is sandwiched between two normal s-wave superconductors\cite{lindner} only when ferromagnetic elements accompany the strain\cite{jianfei} does a $\pi$ shift occur, while as we will see in this work, a strained TI layer sandwiched between two s-wave superconductors leads to a $\pi$-junction without the need of any ferromagnetic elements.  Thus it does not suffer from the limitations of low coherence lengths as d-wave junctions are hampered by nor any extra ferromagnetic element is needed to make the $\pi$ shift possible.  Controlling ferromagnetic elements is an onerous task in such junctions.

\section {Objectives} The specific aim of this letter is to design qubits and gates with strained TI's. We work on an annular ring of topologically insulating material. Specific parts of the ring are rendered superconducting via the proximity effect. Thus, there are two JJ's- one on top the other on the bottom as shown in Fig.~1. Either of them can be tuned to a  $\pi-$phase, by application of a tensile strain (controlled by a simple gate voltage), this creates the necessary double degeneracy for encoding a qubit. By an external magnetic flux one can differentially populate either of these states and thus manipulate the qubit. In particular, the letter would aim to design one-qubit phase  gates. 

\begin{figure} \centerline{\includegraphics[width=11cm,height=7cm]{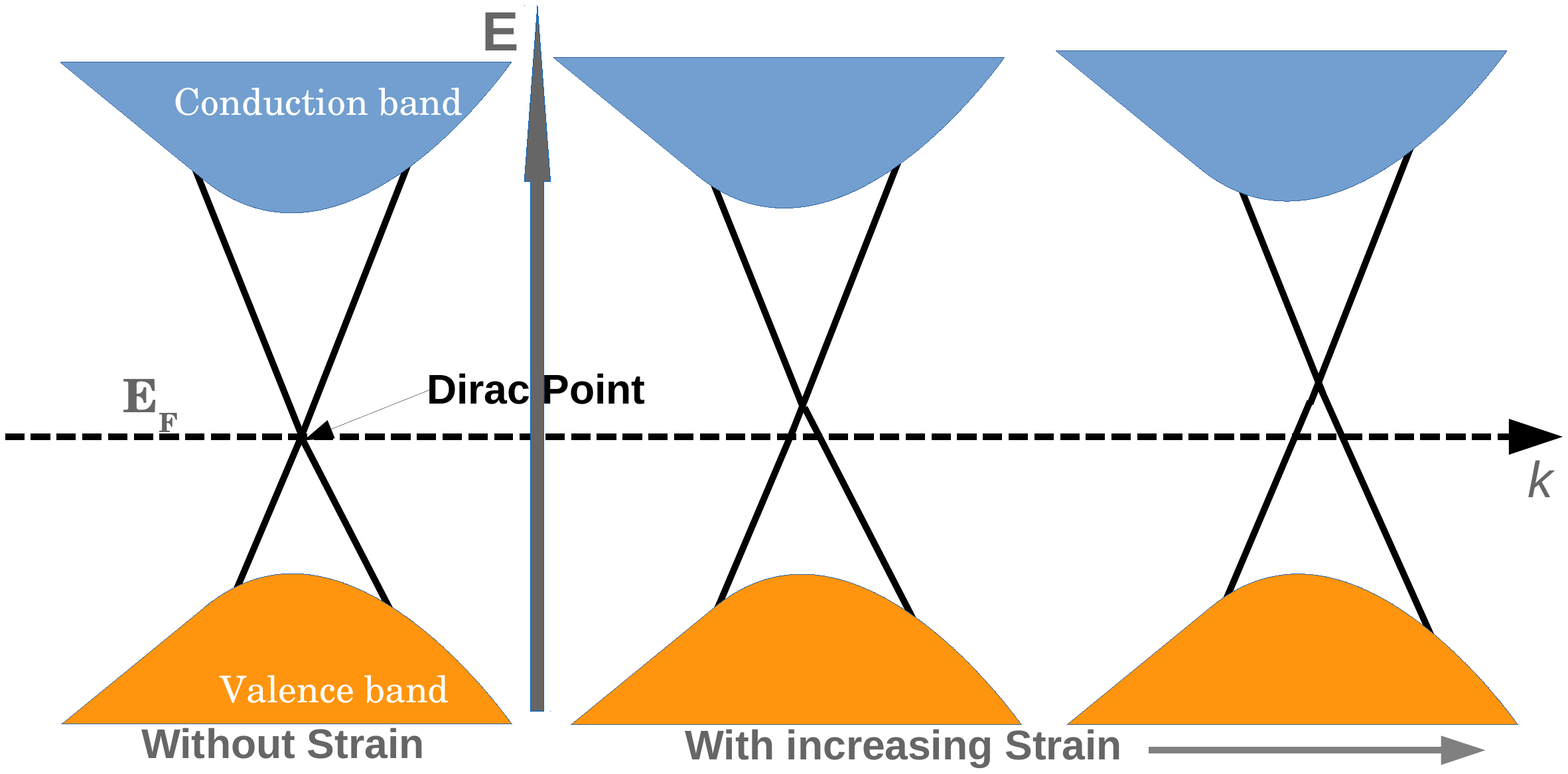}} \vskip -3cm\caption{Shift in the Dirac point of band structure due to the application of a tensile strain.\label{dirac-ti}} \end{figure}
\begin{figure} \centerline{\includegraphics[width=10cm,height=8cm]{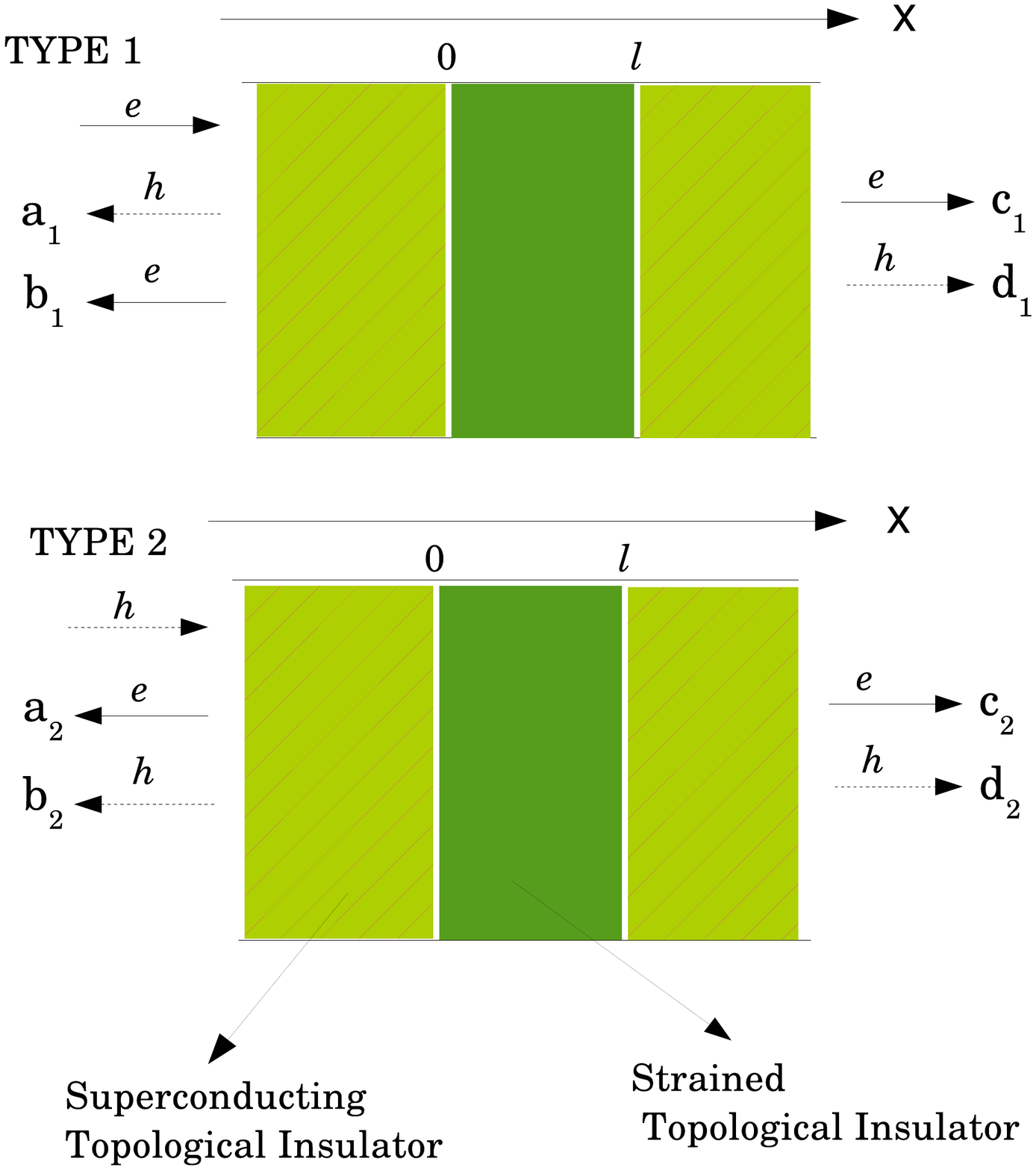}}\vskip -1cm\caption{The Josephson current is calculated by using the Furusaki-Tsukadi approach. This involves calculating the probability amplitudes for electron/hole reflection and transmission for two distinct   processes labeled TYPE 1(Electron quasi-particle incident from the left) and TYPE 2(Hole quasi-particle incident from the left),   see Ref.\cite{fur-tsu,benj-pi}. \label{type}} \end{figure} 

\section{ Theory} The unique Dirac band structure of a TI (as shown in Fig.~2) isn't affected as long as it is a tensile strain. The effect of this tensile strain is captured in Fig.~2. It leads to a shift in the Dirac point without changing the Dirac behavior of the topological surface states (Linear energy versus momenta relationship ). However, a compressive strain does destroy the Dirac nature of the topological surface states, see \cite{jinfeng} for an analysis of the difference between these two kinds of strain. In this letter we confine ourselves to tensile strain only.  Further its been both experimentally observed\cite{jinfeng,yingliu} as well as predicted form DFT electronic structure calculations\cite{jinfeng,zhao,yingliu} that for a tensile strain the shift in Dirac point is proportionally correlated with the magnitude of the strain applied.  Earlier works on JJ in graphene\cite{benj-pi} and molecular magnets\cite{benj-mol} (see references cited therein) have discussed the $\pi$-junction behavior. To see the $\pi$ shift we look at either the top or bottom sections of the ring in Fig. 1 (top panel).  An expanded view of this section is shown in Fig. 3.  The Josephson super-current is measured across the TI surface. A pictorial representation of the processes involved in this is shown in Fig.~3 which depicts the sandwich structure needed to show the $\pi-$phase shift of the Josephson current. 

Quasi-particle scattering in TI's is characterized via the Dirac-Bogoliubov-de Geenes equation\cite{vali}-
\begin{equation}
\left( \begin{array}{cc}
    H  & \Delta(x)\\
    -\Delta^{*}(x)  &-H^{*} \end{array} \right) \Psi(x)e^{i p_{y}y/\hbar v_{F}} = E \Psi(x)e^{i p_{y}y/\hbar v_{F}},
\end{equation}
wherein  $H=v_{F}\vec{p}\cdot\vec{\sigma}-E_{F}+U$. $v_{F}$ denotes Fermi velocity of the quasi particles in TI's and $\vec{\sigma}$ denotes Pauli matrices. The electrostatic potential $U$ is adjusted via gate voltage or doping. $U=0$ in the non-superconducting regions while   $U=-U_{0}$ in the superconducting topological insulators.  The gap $\Delta(x)$ in the  superconducting topological insulators takes the form-
$\Delta(x)=\Delta_{1}e^{i\phi_{1}}\Theta(-x)+\Delta_{2}e^{i\phi_{2}}\Theta(x-l)$. 
\begin{figure*} \centerline{ \includegraphics[width=19cm,height=4cm]{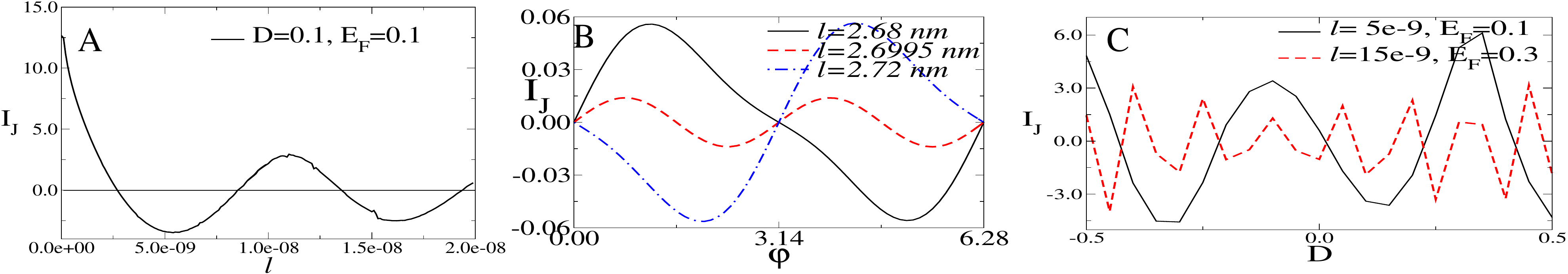}} \caption{A. Josephson supercurrent in units of $e\Delta / \hbar$ versus the width of intervening strained TI layer of length $l$ which varies from zero to $50$nm. Strain($D$) and Fermi energy($E_{F}$) are expressed in eV. B. Josephson supercurrent($I_{J}$) vs the phase for strain ($D=0.1 eV$), Fermi energy ($E_{F}=0.1 eV$) and length of strained region as mentioned in caption . At $l=2.6995 nm$ a $\pi$ periodic Josephson supercurrent is observed. C. Josephson supercurrent versus the strain($D$) in eV. }  \label{J-d} \end{figure*}
The strain is applied to the TI in the region $0<x<l$. In order to design the qubit we have to show the conditions for  $\pi$ junction behavior in the geometry as depicted in Fig.\ref{type}. To do this we have to find the scattering states and from these find the Josephson super-current and Free energy. In Figure below two scenarios are depicted, TYPE 1- wherein an electronic quasi-particle in the state
$\begin{array}{lll}
\Psi^e_{S_{1} +} & = & [u, ue^{i \theta^+}, v, ve^{i
\theta^+}]^T e^{iq^e \cos \theta^+ x},\end{array}$
$q^e = (E_F + U_0 + \sqrt{E^2 - \Delta^2})$ is incident from left superconductor at angle $\theta^{+}$. This incident quasi particle can be reflected as it is, this is described by the substitution $\theta \rightarrow \pi -\theta$. However, it can also be Andreev reflected as a  hole quasi-particle with an angle of reflection
$\theta_{}^-$ and the corresponding wave-function is:
$\begin{array}{lll}
  \Psi^h_{S_{1} -} & = & [v, ve^{- i \theta^-}, u , ue^{- i
\theta^-}]^T e^{- iq^h \cos \theta^- x},
\end{array}$ $q^h = (E_F + U_0 - \sqrt{E^2 - \Delta^2})$.
The superscript e (h) denotes an electron (hole) quasi-particle. Since
translational invariance in the $y$-direction holds the corresponding
component of momentum is conserved. This enables us to calculate the Andreev reflection angle $\theta_A$ through $q^h \sin(\theta_{}^-) = q^e \sin (\theta^+)$. There is no Andreev reflection and
consequently no sub-gap current for angles of incidence above the
critical angle $\theta_c = \sin^{- 1} (|q^h | / q^e)$. The electron/hole coherence factors
are- $u = \sqrt{(1 + \sqrt{1 - \Delta^2 / E^2}) / 2}$, $v =
\sqrt{(1 - \sqrt{1 - \Delta^2 / E^2}) / 2}$. The aforementioned angles are related via $\theta^+
= \theta^e_S,$ and $\theta^- = \pi - \theta^h_S$.

In the strained topological insulator region the eigenvector and corresponding momentum of a right moving electron at an incident angle $\theta$ is
\begin{equation}
  \psi^e_+ = [1, e^{i \theta}, 0, 0]^T e^{ip^e \cos \theta x},
  \hspace{0.25em} \hspace{0.25em} \hspace{0.25em} p^e = [E_{F}+D+E],
\end{equation}

with $D$ being the strength of the strain applied, for a normal topological insulator $D=0$.
A left moving electron is described by the substitution $\theta
\rightarrow \pi - \theta$. If Andreev-reflection takes place, a
left moving hole is generated with energy $E$, angle of reflection
$\theta_A$ and its corresponding wave-function is given by
\begin{equation}
  \psi^h_- = [0, 0, 1, e^{- i \theta_A}]^T e^{- ip^h \cos \theta_A x},
  \hspace{0.25em} \hspace{0.25em} \hspace{0.25em} p^h = [E_{F}+D-E] .
\end{equation}
The transmission angles $\theta^{\alpha}_S$ for the electron and
hole quasi-particles are given by $q^{\alpha} \sin
\theta^{\alpha}_S = p^e \sin \theta, \alpha = e, h$.

The  wave-functions in the three regions as depicted in the TYPE 1 scenario can then be written  as below:
\begin{eqnarray}
  \psi_{S_{1}}&=&\Psi^{e}_{S_{1}+}+b_{1}\Psi^{e}_{S_{1}-}+a_{1}\Psi^{h}_{S_{1}-},
 \,\, x<0,\nonumber\\
  \psi_{N}&=&p\psi^{e}_{+}+q\psi^{e}_{-}+m\psi^{h}_{+}+n\psi^{h}_{-},\,\,0<x<l,\nonumber\\
  \psi_{S_{2}}&=&c_{1}\Psi^{e}_{S_{2}+}+d_{1}\Psi^{h}_{S_{2}+}, \,\, x>l.
\end{eqnarray}
Matching the wave-functions at the interfaces one can solve for the
scattering amplitudes  $a_{1}$, $b_{1}$, $c_{1}$ and $d_{1}$. Similarly,
one can write the wave-functions in case of TYPE 2 scenario and calculate the
amplitudes $a_{2}$, $b_{2}$, $c_{2},$ and $d_{2}$. 

The detailed balance\cite{fur-tsu} for the amplitudes are verified as follows
\begin{eqnarray}
C a_1(\phi,E)&=&C' a_2(-\phi,E),\nonumber\\
b_{i}(\phi,E)&=&b_{i}(-\phi,E) (i=1,2),
\end{eqnarray}
with $C=\sqrt{\frac{\cos\theta^h_S}{\cos\theta^e_S}}$ and
$C'=\sqrt{\frac{\cos\theta^e_S}{\cos\theta^h_S}}$. Following the procedure
established in Ref.~\cite{fur-tsu} and employing analytic continuation $E
\rightarrow iw_n$ the dc Josephson current is given by
\begin{eqnarray}
  I_{J}(\phi)&=&\frac{e\Delta}{2\beta\hbar\Omega_n}\sum_{w_n}\int^{\pi/2}_{-\pi/2}(C+C')
  \Big[\frac{a_{1}(\phi,iw_n)}{C}\nonumber\\
  &-&\frac{a_{2}(\phi,iw_n)}{C'}\Big]\cos(\theta^e_S)d\theta^e_S,
  \label{eq:Ij}
\end{eqnarray}
where $\beta=1/k_{B}T, \Omega_{n}=\sqrt{w^{2}_{n}+\Delta^2}$ and $w_{n}=\pi
k_{B} T (2n+1)$, $n=0, \pm 1, \pm 2, ...$. The above equation has a simple
physical interpretation~\cite{fur-tsu}. Andreev reflection is equivalent to
the breaking up or creation of a Cooper pair. The scattering amplitude $a_1$
describes the process in which an electron quasi-particle coming from
the left superconducting topological insulator ($x<0$) is reflected as a hole
quasi-particle. The amplitude $a_2$ corresponds to the reverse process in
which a hole quasi-particle is reflected as an electron
quasi-particle. This implies that $a_1$ and $a_2$ correspond to the passage
of a Cooper pair to the left and right respectively, hence, the dc Josephson
current is proportional to $a_{1}-a_{2}$. Further, the dc Josephson current
is an odd function of the phase difference, $\phi$, as seen by the detailed
balance condition $a_{2}(\phi,iw_{n})/C=a_{1}(-\phi,iw_{n})/C'$. To
calculate the Josephson current one thus takes the difference between the
amplitudes $a_1$ and $a_2$ and then sums over the energies. The bound state and continuum contributions to the Josephson super-current can also be easily calculated.
Eq.\ref{eq:Ij} can be simplified as-
\begin{eqnarray}
  I_{J}(\phi)&=&\,\,\,\,\sum_{w_n}\frac{e\Delta}{2\beta\hbar\Omega_{n}}\,\,\,\,\int^{\pi/2}_{-\pi/2}
 2iJ \cos(\theta^{e}_{S}) d\theta^e_S, \mbox{ and }\nonumber\\
J&=&\frac{A\sin(\phi)+B\sin(2\phi)}{A'+2B'\cos(\phi)+2C'\cos(2\phi)}
\label{eq:Ij-simp}
\end{eqnarray}
In Eq.\ref{eq:Ij-simp}, $A, B, A', B',$ and $ C'$ are functions of $\theta^{e}_{S}, iw_n, E_F, D \mbox {and } l$.  In the supplementary material the explicit expression for $A, B, A', B',$ and $ C'$ are given. The free energy of the JJ too is calculated as usual from the supercurrent as- \begin{equation} F(\phi)=\frac{1}{2\pi}\int_{0}^{\phi} I_{J}(\phi') d\phi'. \label{eq:Free} \end{equation}
\begin{figure} \centerline{\includegraphics[width=9cm,height=5cm]{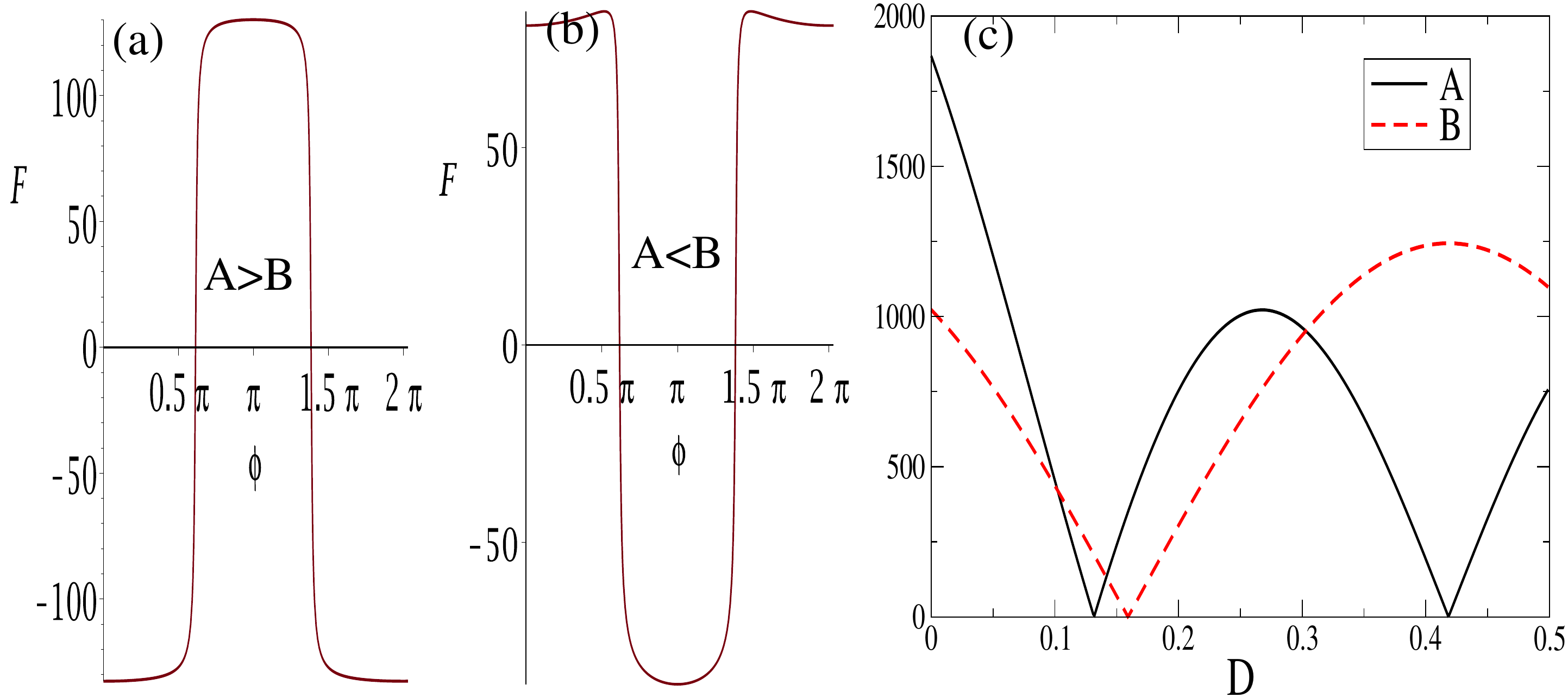}}\caption{Understanding the $\pi$ shift via Eq. 7. In the expression for Josephson current $I_J$, J is the only $\phi$ dependent term. Integrating eq. 7 we have the $\phi$ dependence of the Free energy. Left (a): Free energy curves for $ A=10, B=1$, the $0$ junction. Centre (b): Free energy curves for  $A=1, B=10$, the $\pi$ junction, the other parameters $ A' =2.5, B'=1.41, C'=1$. Right (c): A and B plotted as function of the strain in eV. The other parameters are $E_{F}=0.1 eV, \theta^{e}_{S}=0, E=0.1 meV, \Delta=1 meV$ and length of strained region $1 nm$.} \label{free-maple} \end{figure}

\section {Results} The calculations for the Josephson current as function of the width of the  TI layer as well as the phase difference across the two superconducting topological insulator strips are plotted in Fig. 4. The calculations are performed by treating Eqs.~(\ref{eq:Ij}) and~(\ref{eq:Free}) numerically and the derived results hold for the $T\rightarrow 0$ temperature limit with superconducting gap $\Delta = 1.0 $ meV and $U_{0}=100\Delta=100$ meV so that the Fermi wavelength in the superconductor $1/(E_{F}+U_{0})$ is smaller than the coherence length and Dirac-Bogoliubov-deGennes equations hold. Fig.~\ref{J-d}A  shows the Josephson current as function of the width of the TI layer for different values of tensile strain. The plot shows that for strain values in the range $50-100$meV the Josephson current changes sign, especially in the parameter regime $5 nm < l < 20 nm$, implying a $\pi$ shift.  Fig.~\ref{J-d}B shows the current-phase relation for two different values of strain and the Fermi energy. It again confirms the earlier indication of $\pi$ shift. Now what are the reasons for the occurrence of the $\pi$ junction. From Eq.\ref{eq:Ij-simp} is quite clear that the Josephson current has a periodicity of $\phi=2\pi$ as well as $\pi$. Generally the Josephson supercurrent has $2\pi$ periodicity, however the presence of the $\pi$ term implies depending on the relative magnitude of $A$ or $B$ the $\pi$ or $2\pi$ periodicity will be dominant. In Fig. 5 we plot the integral of the only $\phi$ dependent term of the Josephson supercurrent expression $J$, from Eq.~7. In the limit wherein $A\gg B$ ``dominance of $2\pi$ periodicity'' the system shows a $0$ junction character while in the opposite limit $A \ll B$ ``dominance of $\pi$ periodicity''we have a $\pi$ junction. 	 {In Fig.\ref{free-maple} A and B as function of strain is plotted. It shows the limits wherein $A\gg B$ for $D \sim 0.16 eV$ while  $A\ll B$ for $ D  \sim 0.42 eV$ .}

Further,  any transition from $2\pi$ to $\pi$ periodicity has to accompanied by a phase shift of $\pi$, see Fig.~4(B). This is not unlike what happens in the normal state Aharonov-Bohm effect in metals. There too we see a transition between  flux periodicity of $2\pi$ and $\pi$ accompanied by a phase change of $\pi$ in the Aharonov-Bohm oscillations\cite{lindelof}. The Free energy (via Eq.~8) plotted in Fig.\ref{free-phi}A, plot shows that as one changes the strain via a  gate voltage the ground state of the junction changes from $0$ to $\pi$.

\section {Designing the qubit} 
\begin{figure} \centerline{\includegraphics[width=9cm,height=5cm]{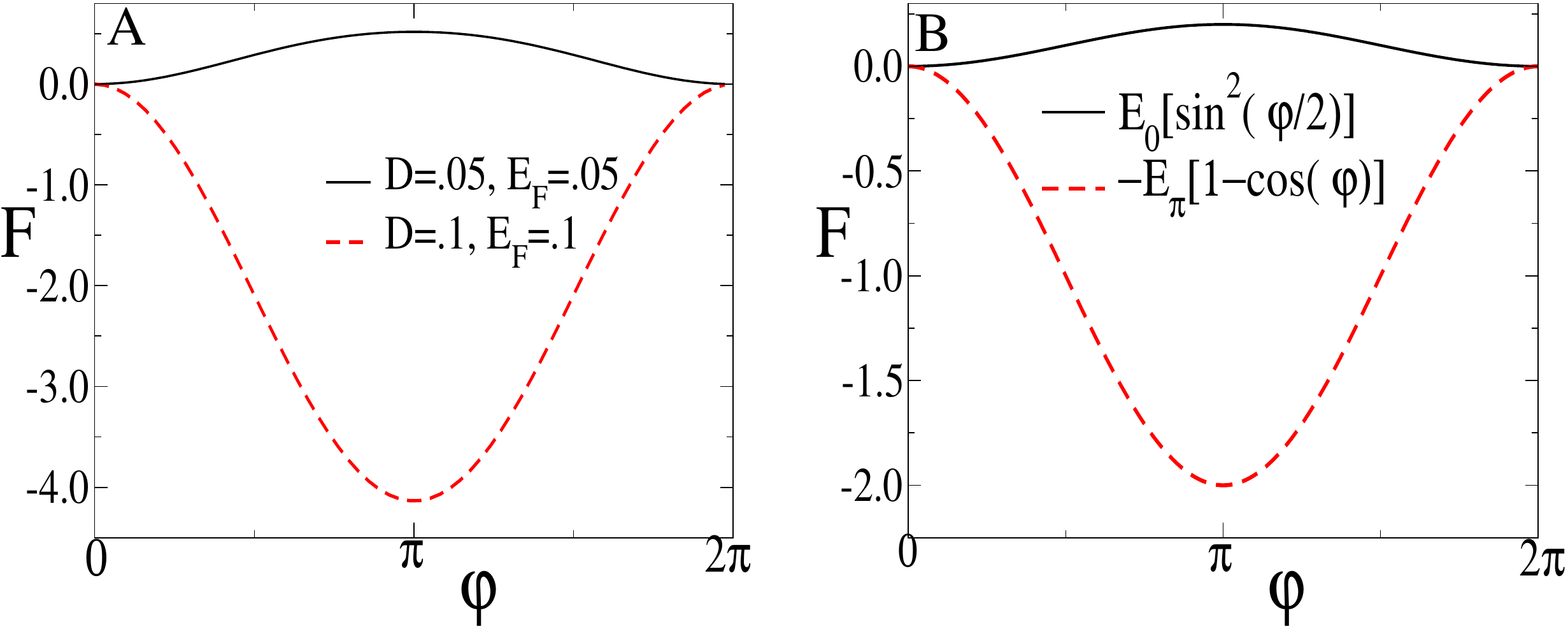}}\caption{Left: Free energy curves for different values of Fermi energy and strain.Right: The approximate forms.} \label{free-phi} \end{figure}

Fig.~\ref{free-phi}A shows the behavior of the Free energy of the JJ. It has a minimum at $\phi=\pi$ (for the $\pi$ junction case) and the variation of F with $\phi$ is strongly dependent on the strain and the Fermi energy. In this parameter regime the free energy can be approximated as $F\sim -E_{\pi} [1-\cos(\phi)]$ as shown in right panel of Fig.~\ref{free-phi}, with $E_{\pi}$ being the Josephson coupling constant. In Fig.~1, the $0$ junction and the $\pi$ junction have Josephson energies $U_{0}=E_{0}[\sin^{2}(\phi/2)]$ and $U_{\pi}=-E_{\pi}[1-\cos(\phi)]$ plotted in Fig.~\ref{free-phi}(Right panel). The superconducting phase difference is $\phi_0$ for the $0$ junction and $\phi_\pi$ for the $\pi$ junction. The total flux in the ring $\Phi$ satisfies $\phi_{\pi}-\phi_{0}=2\pi\Phi/\Phi_0$, where $\Phi_0$ is the flux quantum.

\begin{figure}\centerline{\includegraphics[width=10cm,height=6cm]{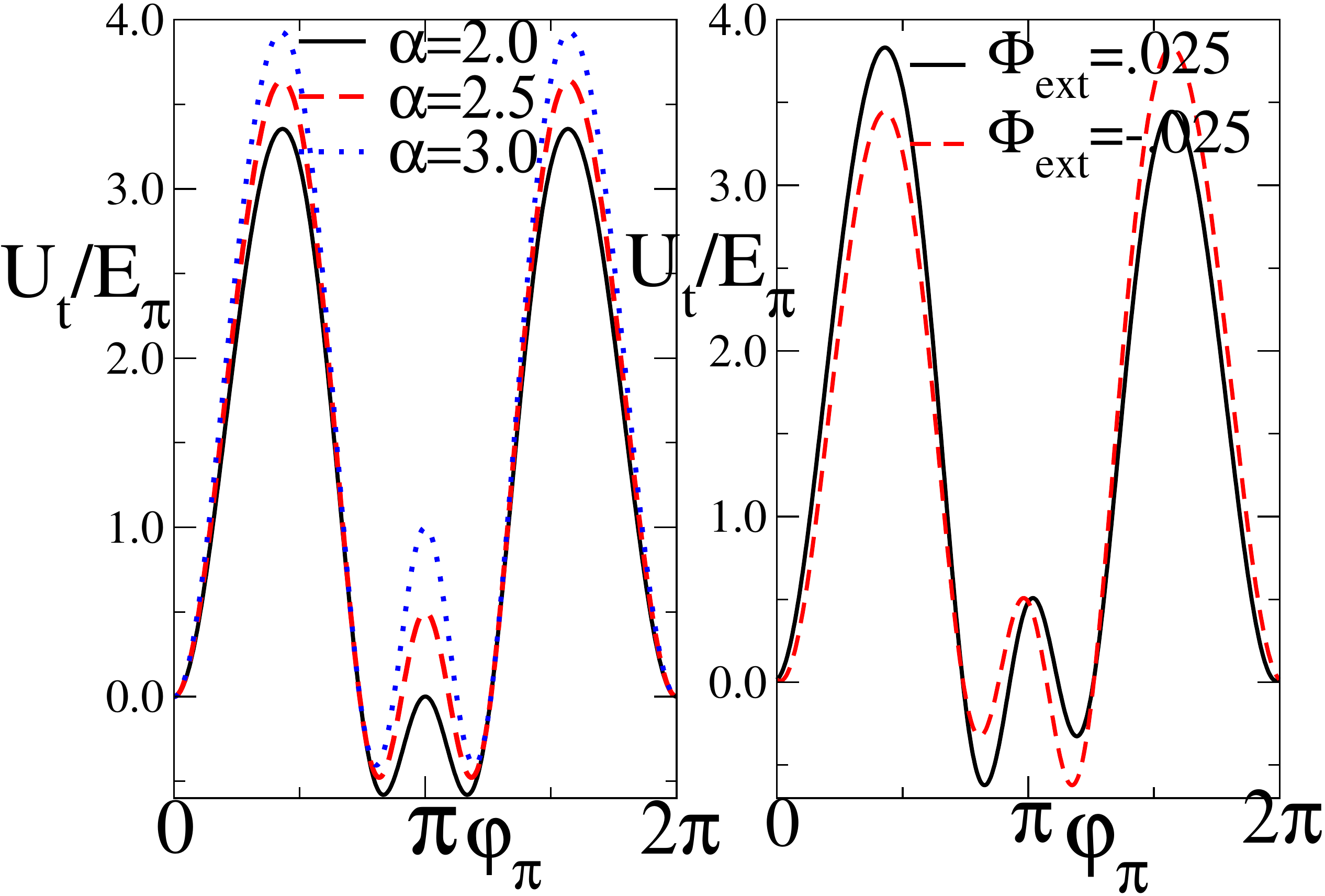}}\caption{Designing the qubit. Left: without any external flux, degenerate ground states at  $\phi_{\pi}\sim 4\pi/5 (|0\rangle$ state) and $6\pi/5 (|\pi\rangle$ state), the qubit basis, for different values of $\alpha=\frac {E_{0}}{E_{\pi}}$ as mentioned and $\beta=0.5$. Right: Finite external field in units of $\Phi_{0}=hc/e$ leads to breaking of the degeneracy. Thus qubit states can be differently populated and any superposition of the basis states can be fashioned.} \label{qubit} \end{figure}

Here we show that single qubit gates can be efficiently implemented in strained Topological insulator JJ's. In Ref.~\cite{maekawa} the authors demonstrate a qubit with a $\pi$ (Superconductor-Ferromagnet-Superconductor) junction and a $0$ (Superconductor-Normal metal-Superconductor) junction coupled into a ring. The qubit dynamics are controlled by an external flux. In our work too we predict that this system, which does not need any ferromagnetic element, could also be prepared similarly in order to implement a qubit. The full Hamiltonian of the TI ring system (Fig. 1) is given by $H=K+U_{t}$ with $U_{t}=U_{0}+U_{\pi}+U_L$, where $U_L=(\Phi-\Phi_{ext})^2/2L_S$ is the magnetic energy stored in the ring, $U_{0}$ and $U_{\pi}$ are the Josephson energies of the $0$ and $\pi$-junctions while $K$ is the flux independent kinetic energy.
We next minimize the Hamiltonian with respect to flux and obtain $\Phi(\phi_\pi)=\Phi_{ext}+\frac{\pi\beta}{2}\sin(\phi)$, with $\beta=2\pi E_{\pi} L_S/\Phi_0^2$. Substituting this equation in the expression for $U_{t}$, we have: 
\begin{eqnarray} U_{t}/E_{\pi}&=&\alpha[\cos^{2}(\phi_{\pi}/2-\frac{\pi\Phi}{\Phi_{0}}-\pi\beta\sin(\phi_{\pi})+\pi/2)]\nonumber\\ &-&[1-\cos(\phi_{\pi})]+2\pi\beta\sin^{2}(\phi_{\pi}). \label{minH} 
\end{eqnarray} with $\alpha=E_0 / E_\pi$. For typical values mentioned in Fig.~\ref{qubit}, we plot Eq.~(\ref{minH}). The energy of of the TI ring system (Fig.~\ref{qubit}) shows a double minima located approximately  at $\phi_{\pi}\sim 4\pi/5 (|0\rangle$ state) and $6\pi/5 (|1\rangle$ state), this doubly degenerate state is the basis of the qubit. The state can tunnel between these wells depending on value of barrier around $\phi_{\pi} \sim \pi$. The strength of barrier hence tunneling between $|0\rangle$  and $|1\rangle$ states is controlled via $\alpha=\frac{E_{0}}{E_{\pi}}$, as $\alpha$ increases the barrier increases and tunneling will hence decrease. Thus we can go from an equal superposition to either a $|0\rangle$ or $|1\rangle$ state by tuning $\alpha$. As we have already seen in Fig.~\ref{qubit} , ${E_{0}}$ and ${E_{\pi}}$ are dependent on  strain and Fermi energy. These quantities can be very easily controlled in a TI thus making the qubit easily tunable via strain and/or Fermi energy. The right panel of Fig.~\ref{qubit} shows an external control brought about via a magnetic field. A finite external field breaks the degeneracy and leads to two qubits states being differently populated meaning any superposition of qubit basis states can be achieved. The simplest one qubit gate is the phase gate, this takes a qubit from an initial state to a final state which differs from the initial by a phase factor. One can estimate the time taken to implement a basic $\pi$ phase gate which takes a qubit for example from $|+\rangle$ to $|-\rangle$ state is $\tau_{phase}=\pi\hbar/\delta E$ wherein $\delta E$ is the energy difference effectively between $E_{\pi}$ and $E_{0}$. The junction energy $E_{0}$ in our case is defined as the product of critical current flowing in a TI based Josephson junction(JJ)  and the flux quantum $\hbar/2e$ which gives $10^{-20}$ Joules for a critical current of $20 \mu A$ flowing in a typical TI based JJ\cite{veldhorst}. As $\alpha =\frac{E_{0}}{E_{\pi}}=3.0$, $\delta E\sim 10^{-20}$ Joules. Therefore time taken to implement a phase gate is $\tau_{phase}=1$ pico seconds which is much less than the decoherence time of Josephson flux qubits which is around microseconds\cite{zagoskin}.

 The DiVincenzo criteria is related to the necessities for the design of a quantum computer which has many many qubits and gates\cite{kais}. In this work we confine ourselves only to the design of a single qubit and by extension to a single qubit gate-the phase gate. The DiVincenzo criteria which relate to a single qubit are\cite{kais}:  1. The ability to initialize the state of a qubit to a simple fiducial state and  related to gates, 2. relatively long decoherence times. In the letter we have shown how by the gate voltage controlled strain and an external magnetic flux we can switch between the states of the qubit, i.e., if initially it is in `0' state we can swith to `1' or viceversa. We can also rotate the state of the qubit by making any arbitrary superposition of `0' and `1' possible and thus implement a phase gate. Further the operating time of the phase gate is  in pico seconds, much less than the decoherence time of Josephson flux qubits. 

Finally what in our proposal is new- the proposal aims to design a $\pi$-junction qubit in a topological insulator which has never been attempted in this material.  A $\pi$ Josephson junction\cite{golubov}  till date has never been observed without having either one of the following (1) Ferromagnets (2) A magnetic impurity/magnetic molecule/Kondo impurity in  a Josephson junction or (3) A unconventional order parameter (e.g., d-wave/p-wave), in any material let alone Dirac materials like graphene or topological insulators. This is the first time we show the design of a $\pi$ Josephson junction without using any of the above three ingredients. We have normal s-wave superconducting correlations on either side of the strained topological insulator. We have a tensile strain which generates the $\pi$ shift.  This is the unique result of our work.  After this we show how this $\pi$-junction can be utilized to design a qubit. The qubit thus is uniquely generated and controlled via the applied tensile strain. 
In conclusion we provide a perspective on future endeavors in this area.  We plan to take this forward and design  different kinds of single and two qubit gates using a strained layer of TI sandwiched in between two superconducting layers. Further, another fruitful extension would be to check on existence of Majorana fermions at the interface of a TI and superconductor\cite{zhang,benj-majo} by the use of  strain.  
\section{ Acknowledgments} This work was supported by funds from Dept. of Science and Technology (Nanomission), Govt. of India, Grant No. SR/NM/NS-1101/2011.

\begin{widetext}
\section{Supplementary Material}
The exact expression for Josephson supercurrent as mentioned in Eq. 7 of the main text. The explicit expressions for the coefficients $A, B, A', B' \mbox{and} C'$ are mentioned.
 \begin{eqnarray}
      J &=&2i \frac{A \sin(\phi)+B \sin(2\phi)}{A'+2 B' \cos(\phi)+2 C' \cos(2 \phi)},\\
\mbox{with } A &=& y1\cdot x145\cdot f  h+y23\cdot x2\cdot f^{2}-y23\cdot x3\cdot f^{2}-y4\cdot x145\cdot f g\nonumber\\
 B &=& (y1\cdot x2\cdot h-y4\cdot x3\cdot g)\cdot f^{3}\nonumber\\
      A' &=&x145\cdot x145+(x2\cdot x2+x3\cdot x3)\cdot f^{4}\nonumber\\
      B' &=&(x2+x3)\cdot x145\cdot f^{2}\nonumber\\
      C' &=&x2\cdot x3\cdot f^{4}\nonumber\\
 \mbox{where, }  y1 &=& (xn2\cdot xm1-xm2\cdot xn1)\cdot (xq2p\cdot xp1p-xq1p\cdot xp2p),\nonumber\\
      x1 &=& (xp3\cdot xq2p-xq3\cdot xp2p)\cdot (xn1p\cdot xm2-xm1p\cdot xn2),\nonumber\\
x4 &=& (xn1p\cdot xm3-xm1p\cdot xn3)\cdot (xq2\cdot xp2p-xp2\cdot xq2p),\nonumber\\
    &+&(xq1p\cdot xp3-xq3\cdot xp1p)\cdot (xn2\cdot xm2p-xm2\cdot xn2p),\nonumber\\
   x5 &=& (xp1p\cdot xq2-xp2\cdot xq1p)\cdot (xn3\cdot xm2p-xm3\cdot xn2p),\nonumber\\
   x145 &=& x1\cdot h^{2}+x4\cdot f^{2}+x5\cdot g^{2},\nonumber\\
 x2 &=& (xq2\cdot xp3-xp2\cdot xq3)\cdot (xm1p\cdot xn2p-xn1p\cdot xm2p),\nonumber\\
  x3 &=& (xq1p\cdot xp2p-xp1p\cdot xq2p)\cdot (xn3\cdot xm2-xm3\cdot xn2),\nonumber\\
y2 &=& (xq2\cdot xp2p-xp2\cdot xq2p)\cdot (xn1p\cdot xm1-xm1p\cdot xn1)\nonumber\\
      &+&(xm1p\cdot xn2-xn1p\cdot xm2)\cdot (xq1\cdot xp2p-xq2p\cdot xp1),\nonumber\\
     y3 &=& (xn2p\cdot xm1-xn1\cdot xm2p)\cdot (xp2\cdot xq1p-xq2\cdot xp1p)\nonumber\\
        &+&(xq1p\cdot xp1-xq1\cdot xp1p)\cdot (xm2p\cdot xn2-xm2\cdot xn2p),\nonumber\\
 y23 &=& (y2\cdot h+y3\cdot g)\cdot f\nonumber,\\
 y4 &=& (xp2\cdot xq1-xq2\cdot xp1)\cdot (xn1p\cdot xm2p-xm1p\cdot xn2p),\nonumber\\
     xp1 &=& (\exp(i \theta_{S}^{e})+\exp(-i \theta))/(2 \cos(\theta)),\nonumber\\
      xp2 &=& (\exp(-i \theta)-\exp(-i \theta_{S}^{e}))/(2 \cos(\theta)),\nonumber\\
      xp3 &=& (\exp(-i \theta)+\exp(i \theta_{S}^{h}))/(2 \cos(\theta)),\nonumber\\
xp1p &=& ((\exp(-i \theta)+\exp(i\theta_{S}^{e}))/(2 \cos(\theta))) \exp(-i  p^{e} \cos(\theta) l)\nonumber\\
xp2p &=& ((\exp(-i\theta)-\exp(-i\theta_{S}^{h}))/(2 \cos(\theta))) \exp(-i  p^{e} \cos(\theta) l)\nonumber\\
      xq1 &=& ((\exp(i \theta)-\exp(i \theta_{S}^{e}))/(2 \cos(\theta)))\exp(-i p^{e} \cos(\theta) l)\nonumber\\
      xq2 &=& ((\exp(i \theta)+\exp(-i \theta_{S}^{e}))/(2 \cos(\theta))) \exp(-i p^{e} \cos(\theta) l)\nonumber\\ 
      xq3 &=& ((\exp(i \theta)-\exp(i \theta_{S}^{h}))/(2 \cos(\theta))) \exp(-i p^{e} \cos(\theta) l)\nonumber\\
      xq1p &=& (\exp(i \theta)-\exp(i \theta_{S}^{e}))/(2 \cos(\theta))\nonumber\\
      xq2p &=& (\exp(i \theta)+\exp(-i \theta_{S}^{h}))/(2\cos(\theta))\nonumber\\
      xn1 &=& ((\exp(i \theta_{A})+\exp(i \theta_{S}^{e}))/(2 \cos(\theta_{A}))) \exp(-i p^{h} \cos(\theta_{A}) l)\nonumber\\
      xn2 &=& ((\exp(i\theta_{A})-\exp(-i \theta_{S}^{e}))/(2 \cos(\theta_{A}))) \exp(-i p^{h} \cos(\theta_{A}) l)\nonumber\\
      xn3 &=& ((\exp(i \theta_{A})+\exp(i \theta_{S}^{h}))/(2\cos(\theta_{A}))) \exp(-i p^{h}\cos(\theta_{A})l)\nonumber\\  
      xn1p &=& (\exp(i \theta_{A})+\exp(i \theta_{S}^{e}))/(2 \cos(\theta_{A}))\nonumber\\
      xn2p &=& (\exp(i \theta_{A})-\exp(-i \theta_{S}^{h}))/(2 \cos(\theta_{A}))\nonumber\\
      xm1 &=& (\exp(-i \theta_{A})-\exp(i \theta_{S}^{e}))/(2 \cos(\theta_{A}))\nonumber\\
      xm2 &=& (\exp(-i \theta_{A})+\exp(-i \theta_{S}^{e}))/(2 \cos(\theta_{A}))\nonumber\\
      xm3 &=& (\exp(-i \theta_{A})-\exp(i \theta_{S}^{h}))/(2 \cos(\theta_{A}))\nonumber\\
      xm1p&=&((\exp(-i \theta_{A})-\exp(i\theta_{S}^{e}))/(2 \cos(\theta_{A}))) \exp(-i p^{h} \cos(\theta_{A})l)\nonumber\\
      xm2p&=&((\exp(-i \theta_{A})+\exp(-i \theta_{S}^{h}))/(2 \cos(\theta_{A})))\exp(-i p^{h} \cos(\theta_{A}) l)\nonumber
 \end{eqnarray}
 \end{widetext}
 \newpage
\begin{widetext}
\vspace{-4cm}
  \begin{eqnarray}
  \mbox{with }       q^{e}&=&(E_{F}+U_{0}+\sqrt{\Delta^{2}-E^{2}}))/(\hbar v_{F}),      \nonumber\\
      q^{h}&=&(E_{F}+U_{0}-\sqrt{\Delta^{2}-E^{2}}))/(\hbar v_{F}),\nonumber\\
      x&=&\sqrt{\Delta^{2}-E^{2}}/E,  f=\Delta/(2 E),  g=(1+x)/2,h=(1-x)/2,      \nonumber\\
     p^{e}&=&(E_{F}+E+D)/(\hbar v_{F}), \nonumber\\
       p^{h}&=&(E_{F}-E+D)/(\hbar v_{F}).\nonumber
 \end{eqnarray}                            
   The values of the variables appearing in the above equation are: $  \Delta=1 \mbox{meV},    k_{b}=8.6 * 10^{-5} eV/K, \hbar=6.6*10^{-16} eV. sec,  v_{F}=10^{6} m s^{-1}.$\\
 \end{widetext}
\end{document}